\newcommand{\reff}[1]{(\ref{#1})}
\newcommand{\of}[1]{\left( #1 \right)}

\newcommand{\lmax}{\ensuremath{\lambda_{\text{max}}}}
\newcommand{\vmax}{\ensuremath{\mathbf{v}_{\text{max}}}}
\newcommand{\tr}{\ensuremath{\text{tr}}}

\newcommand{\Rs}{\ensuremath{R_{\Sigma}}}
\newcommand{\RsT}{\ensuremath{R_{\Sigma,\text{TDMA}}}}
\newcommand{\RsTinf}{\ensuremath{R_{\Sigma,\text{TDMA},\infty}}}
\newcommand{\Rsupo}{\ensuremath{R_{\Sigma,\text{up,1}}}}
\newcommand{\Rsupt}{\ensuremath{R_{\Sigma,\text{up,2}}}}
\newcommand{\Ro}{\ensuremath{R^{(1)}}}
\newcommand{\Rk}{\ensuremath{R^{(k)}}}
\newcommand{\Rkinf}{\ensuremath{R^{(k)}_\infty}}
\newcommand{\Ri}{\ensuremath{R^{(i)}}}
\newcommand{\Rj}{\ensuremath{R^{(j)}}}

\newcommand{\Pk}{\ensuremath{P^{(k)}}}
\newcommand{\Pj}{\ensuremath{P^{(j)}}}
\newcommand{\Po}{\ensuremath{P^{(1)}}}
\newcommand{\Pmax}{\ensuremath{P_{\text{max}}}}

\newcommand{\A}{\ensuremath{\mathbf{A}}}
\newcommand{\F}{\ensuremath{\mathbf{F}}}
\newcommand{\Fo}{\ensuremath{\mathbf{F}^{(1)}}}
\newcommand{\Fk}{\ensuremath{\mathbf{F}^{(k)}}}
\newcommand{\R}{\ensuremath{\mathbf{R}}}
\newcommand{\T}{\ensuremath{\mathbf{T}}}
\newcommand{\W}{\ensuremath{\mathbf{W}}}
\newcommand{\I}{\ensuremath{\mathbf{I}}}

\newcommand{\hkr}{\ensuremath{\mathbf{h}^{(k)}_r}}
\newcommand{\hjr}{\ensuremath{\mathbf{h}^{(j)}_r}}
\newcommand{\hro}{\ensuremath{\mathbf{h}^{(1)}_r}}
\newcommand{\hkd}{\ensuremath{h^{(k)}_d}}
\newcommand{\hjd}{\ensuremath{h^{(j)}_d}}
\newcommand{\hdo}{\ensuremath{h^{(1)}_d}}
\newcommand{\hkeff}{\ensuremath{\mathbf{h}^{(k)}_\text{eff}}}

\newcommand{\rmo}{\ensuremath{r^{-1}}}
\newcommand{\rmh}{\ensuremath{r^{-1/2}}}

\newcommand{\btau}{\ensuremath{\boldsymbol{\tau}}}
\newcommand{\btaus}{\ensuremath{\boldsymbol{\tau}^{\ast}}}
\newcommand{\nus}{\ensuremath{\nu^{\ast}}}
\newcommand{\tauk}{\ensuremath{\tau^{(k)}}}
\newcommand{\taukstar}{\ensuremath{{\tau^{(k)}}^\ast}}
\newcommand{\tauj}{\ensuremath{\tau^{(j)}}}
\newcommand{\taujstar}{\ensuremath{{\tau^{(j)}}^\ast}}
\newcommand{\taui}{\ensuremath{\tau^{(i)}}}
\newcommand{\tauistar}{\ensuremath{{\tau^{(i)}}^\ast}}

\newcommand{\tauks}{\ensuremath{\tau^{(1)},\ldots,\tau^{(K)}}}

\newcommand{\xk}{\ensuremath{x^{(k)}}}
\newcommand{\xr}{\ensuremath{\mathbf{x}_r}}
\newcommand{\yr}{\ensuremath{\mathbf{y}_r}}
\newcommand{\zr}{\ensuremath{\mathbf{z}_r}}
\newcommand{\hh}{\ensuremath{\mathbf{h}^H}}
\newcommand{\h}{\ensuremath{\mathbf{h}}}
\newcommand{\y}{\ensuremath{\mathbf{y}}}
\newcommand{\yt}{\ensuremath{\mathbf{\widetilde{y}}}}
\newcommand{\zt}{\ensuremath{\mathbf{\widetilde{z}}}}

\newtheorem{theo}{Theorem}

\documentclass[conference,a4paper]{IEEEtran}

\ifCLASSINFOpdf
  \usepackage[pdftex]{graphicx}
  \graphicspath{{Figures/}}
  \DeclareGraphicsExtensions{.pdf,.jpeg,.png}
\else
  \usepackage[dvips]{graphicx}
  \graphicspath{{Figures/}}
  \DeclareGraphicsExtensions{.eps}
\fi

\usepackage{url}
\usepackage[pdfpagelabels]{hyperref}

\usepackage{cite}
\usepackage{amsmath, amsfonts}
\usepackage{pgf}
\usepackage{tikz}
\usepackage{pgfplots}
\usetikzlibrary{plotmarks}

\usepackage{algpseudocode}
\usepackage{algorithm}

\usepackage{subfigure}

\newlength{\myarraycolsep}
\newlength{\oldarraycolsep}
\renewcommand{\det}[1]{\left| #1\right|}
\newcommand{\norm}[1]{\ensuremath{\left\| #1 \right\|}}

\begin{document}
\title{Achievable Sum-Rates in Gaussian Multiple-Access Channels with MIMO-AF-Relay and Direct Links}

\author{\IEEEauthorblockN{Frederic Knabe, Omar Mohamed, and Carolin Huppert}
\IEEEauthorblockA{Institute of Communications Engineering\\
Ulm University, Albert-Einstein-Allee 43, 89081 Ulm, Germany\\
Email: \{frederic.knabe, omar.mohamed, carolin.huppert\}@uni-ulm.de}
}

\maketitle

\begin{abstract}
We consider a single-antenna Gaussian multiple-access channel (MAC) with a multiple-antenna amplify-and-forward (AF) relay, where, contrary to many previous works, also the direct links between transmitters and receiver are taken into account. For this channel, we investigate two transmit schemes: Sending and relaying all signals jointly or using a time-division multiple-access (TDMA) structure, where only one transmitter uses the channel at a time. While the optimal relaying matrices and time slot durations are found for the latter scheme, we provide upper and lower bounds on the achievable sum-rate for the former one. These bounds are evaluated by Monte Carlo simulations, where it turns out that they are very close to each other. Moreover, these bounds are compared to the sum-rates achieved by the TDMA scheme. For the asymptotic case of high available transmit power at the relay, an analytic expression is given, which allows to determine the superior scheme.
\end{abstract}


% \fi
\IEEEpeerreviewmaketitle

\section{Introduction}
In today's wireless communication systems, the demand for higher data rates and wide-range coverage is steadily growing. To meet these requirements, a high density of base stations is necessary, which entails high costs for installation and maintenance. Another possibility to increase throughput and coverage is the use of relay nodes, which have much lower costs. Relay channels were considered in \cite{CG79} first, and have drawn more and more research attention in the last decades.

Depending on how the signals are processed at the relay, different types of relaying schemes are distinguished. The most common ones are amplify-and-forward (AF, also called non-regenerative relaying) and decode-and-forward (DF, also called regenerative relaying). While in AF, the relay simply amplifies the received signals subject to a power constraint, a complete decoding and re-encoding of the signal is necessary when using DF. As this yields higher costs and larger delays, we will restrict ourselves to AF relaying schemes in this paper.

For multiple input multiple output (MIMO) systems with additive white Gaussian noise (AWGN), the main challenge is to find both the covariance matrix at the transmitter and the matrix that maps the relay's input to its outputs, such that the data rate is maximized. The problem becomes even harder to solve, if a relay system with multiple transmitters, also called a multiple-access relay channel (MARC), is considered. This holds especially if the direct links between transmitters and receiver are also taken into account. A solution for this general problem has not been found yet. However, numerous previous works have made considerable progress at least for some simplified versions of the problem:

If the direct links are neglected, the optimal structure of both the relaying matrix and the transmit covariance matrices has been found \cite{FHK06, YH10}. However, this structure still contains parameters that are subject to optimization and the optimal solution of this problem remains unknown. For the case of a single receive antenna, the above problem could be solved in \cite{KMH12}, where it was also shown that time-division multiple-access (TDMA) further increases the achievable sum-rate. In \cite{TH07}, a single-user system was considered, where the transmit covariance matrices were fixed to scaled identity matrices. With this restriction, an algorithm was found that optimizes the relay matrix. However, for the case of non-zero direct links, only upper and lower bounds could be provided. Different from all previously mentioned works, a half-duplex relay was assumed in \cite{VSBS06}. This relay was used in single-user systems both with and without direct links, for which suboptimal transmit strategies based on iterative algorithms were derived.

In this work, we consider a full-duplex $K$-user MARC with an AF-relay and non-zero direct links, where only the relay has multiple antennas. For this system, we first derive new upper and lower bounds on the achievable sum-rate for the case where all transmitters send their signals jointly. Subsequently, we will extend the TDMA-based transmission scheme introduced in \cite{KMH12} to the case where the direct links are present. An optimal solution for this scheme is achieved by an iterative algorithm. Finally, the achievable sum-rates of the TDMA and the ``joint relaying'' scheme will be compared, where it can be seen that the superiority of TDMA found in \cite{KMH12} does not always persist for the case of non-zero direct links.

This paper is structured as follows: In Section \ref{sec:problem_formulation}, we introduce the channel model and describe the constraints that have to be fulfilled while optimizing the sum-rate. Subsequently, we derive upper and lower bounds for the joint relaying scheme in Section \ref{sec:joint_relaying}. The TDMA scheme is discussed in Section \ref{sec:TDMA}, where also an algorithm achieving the optimal solution and an asymptotic comparison to joint relaying is given. Further comparisons for more general scenarios are provided in Section \ref{sec:simres} by means of simulation results. Finally, Section \ref{sec:conclusion} concludes the paper.

\section{Channel model}\label{sec:problem_formulation}

\subsection{Notation}
We denote all column vectors in bold lower case and matrices in bold upper case letters. The trace, determinant, Hermitian, and transpose of a matrix $\A$ are identified by $\tr(\A)$, $\det{\A}$, $\A^H$, and $\mathbf{A}^T$, respectively. We use $\lVert \mathbf{x} \rVert$ to denote the Euclidean norm of a vector $\mathbf{x}$ and  $\I$ to describe the identity matrix. Furthermore, $\lmax(\A)$ and $\vmax(\A)$ indicate the largest eigenvalue of a matrix $\A$ and its corresponding eigenvector.

\subsection{Channel Model}\label{subsec:channel_mod}

\begin{figure}
\centering{\includegraphics[width=\linewidth]{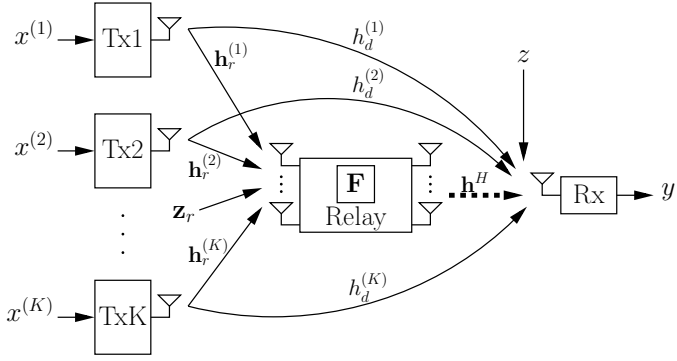}}
\caption{$K$-user multiple-access relay channel (MARC)}
\label{fig:MARC}
\vspace{-0.5cm}
\end{figure}
The MARC that we consider is depicted in Figure \ref{fig:MARC} and consists of $K$ transmitting nodes. Each user $k\in \{1,\ldots,K\}$ transmits the signal $\xk \in \mathbb{C}$, which reaches both the relay and the receiver. The channel matrix for the  transmission to the relay is given by the vector $\hkr \in \mathbb{C}^{M_r}$, while the channel to the receiver is described by the scalar $\hkd \in \mathbb{C}$. Thus, the received signal $\yr$ of the relay can be written as
\begin{equation*}
\yr = \sum\limits_{k=1}^{K} \hkr \xk + \zr,
\end{equation*}
where $\zr \sim \mathcal{CN}(0,\I)$ is the additive white Gaussian noise at the relay and $M_r$ denotes the number of antennas at the relay. The relay amplifies the signals by the matrix $\F$ and transmits the signal $\xr = \F\yr$ over the channel $\hh \in \mathbb{C}^{1\times M_r}$ to the receiver. 

It is assumed that the transmission from the relay to the receiver takes place in a different frequency band, i.e., the signals transmitted by the relay are orthogonal to the signals transmitted by the users. To overcome the problem that the signals from the relay arrive with the delay of one symbol, we assume that the direct signal can be buffered. Hence, the out-of-band reception can be modeled by a virtual second antenna at the receiver, which receives only the signal from the relay. This results in the received signal
\begin{equation*}
\y = \begin{bmatrix}
\hh\F\sum\limits_{k=1}^{K} \hkr \xk + \hh\F\zr + z_1\\
\sum\limits_{k=1}^{K} \hkd \xk + z_2
\end{bmatrix},
\end{equation*}
where $z_i \sim \mathcal{CN}(0,N_0)$ ($i=1,2$) denote the Gaussian noise terms at the receiver\footnote{Throughout this paper, we assume $N_0=1$}. As the first component of $\y$ contains noise both from the relay and from the receiver, it can be simplified by normalization without changing the systems properties. Thus, we will use
\begin{equation}
\label{eq:MARC}
\yt = \sum\limits_{k=1}^{K} \hkeff \xk + \zt
\end{equation}
as equivalent output at the receiver, where $\zt \sim \mathcal{CN}(0,N_0\I)$, $\hkeff = \left[\rmh\hh\F\hkr,\hkd\right]^T$, and $r=1 + \hh\F\F^H\h$.

Both the transmitters and the relay are subject to average power constraints, which are given by
\begin{align}
\label{eq:PowerConstr}
E\of{\norm{\xk}^2} \leq \Pk \nonumber \;\; \forall k\in \{1,\ldots,K\} \\
E\!\of{\tr\of{\xr\xr^H}} \!=\! \tr\of{\!\F\!\of{\I+\sum\limits_{k=1}^{K} {\hkr\Pk\hkr}^H}\!\F^H\!} &\!\leq\! P_r. 
\end{align}
Finally, we assume that perfect channel state information is available at all nodes.

\section{Joint Relaying Scheme}\label{sec:joint_relaying}
The transmit strategy, where all transmitters send their signals at the same time will be referred to as ``joint relaying'' in the remainder of this paper. In this case the MARC from (\ref{eq:MARC}) can be interpreted as a pure single input multiple output (SIMO) multiple-access channel (MAC). For this SIMO MAC, the achievable sum-rate can be optimized by influencing the channel gain through the choice of the relaying matrix $\F$. As in the MAC, the sum-rate can be calculated as
\begin{equation*}
\Rs = \log_2 \det{\I + \sum\limits_{k=1}^{K} \hkeff\Pk{\hkeff}^H}.
\end{equation*}
Evaluating this determinant by using the definition of $\hkeff$ and standard transformations of linear algebra, $\Rs$ can be reformulated as
\begin{align}
\Rs &= \log_2 \of{1 + s + \rmo\hh\F\of{\of{1+s}\R - \T}\F^H\h} \label{eq:Rs_pre}\\
&= \log_2 \of{1 + s + \rmo\hh\F\of{\R + \W}\F^H\h},\label{eq:Rs}
\end{align}
where $s = \sum\limits_{i=1}^{K}\norm{\hkd}^2 \Pk$, $\R = \sum_{k=1}^{K}\hkr\Pk{\hkr}^H$, and
\begin{align*}
\T &= \of{\sum\limits_{k=1}^{K}\hkr\Pk\hkd} \of{\sum\limits_{k=1}^{K}{\hkr}^H\Pk{\hkd}^H}\\
\W &= \frac{1}{2}\!\!\sum\limits_{j,k=1}^{K}\!\!\!\!\of{\!\hjd\hjr \!\!-\! \hkd\hkr\!}\!\!\of{\!\hjd\hjr \!\!-\! \hkd\hkr\!}^H \!\!\!\Pj\!\Pk\!.
\end{align*}
As it can be seen from the above equations the choice of $\F$ only influences the last term inside the logarithm in \reff{eq:Rs_pre} and \reff{eq:Rs}, while the other terms are constant. However, compared to the optimization problem with absent direct links \cite{KMH12}, we have the additional term $\W$, which occurs in the sum-rate but not in the power constraint \reff{eq:PowerConstr}. Hence, the optimal relaying matrix is not the same as in \cite{KMH12}. As the optimal solution seems to be hard to find, we will derive upper and lower bounds in the following two subsections.

\subsection{Upper Bounds on $\Rs$}
A first upper bound can be obtained from the fact that $\T$ is a positive semidefinite matrix. Thus, $\Rs$ can be upper bounded by
\begin{equation}
\label{eq:Rsup1_pre}
\Rs \leq \log_2 \of{1 + s + \rmo\hh\F\of{\of{1+s}\R}\F^H\h}.
\end{equation}
Besides the additive term $s$ and the constant factor $(1+s)$ the maximization problem is now similar to the one in \cite{KMH12}, such that this upper bound is optimized by choosing (cf. \cite{FHK06,YH10})
\begin{equation}
\label{eq:F_ub1}
\F = \sqrt{\frac{P_r}{1 + \lmax(\R)}}\cdot\frac{\h}{\norm{\h}}\cdot\vmax^H(\R).
\end{equation}
Using this relaying matrix in the right side of \reff{eq:Rsup1_pre}, we obtain
\begin{equation*}
\Rsupo = \log_2 \of{(1+s)\of{1+\frac{\lmax(\R)\norm{\h}^2 P_r}{1 + \norm{\h}^2 P_r + \lmax(\R)}}}.
\end{equation*}
A second upper bound can be obtained by ignoring the relay power constraint, i.e., by letting $P_r \rightarrow \infty$, which delivers
\begin{align}
\Rs &\leq \log_2 \of{1 + s + \frac{\hh\F\of{\R + \W}\F^H\h}{\hh\F\F^H\h}}\notag\\
&\leq \log_2 (1 + s + \lmax(\R + \W)) = \Rsupt \label{eq:Rsup2},
\end{align}
where the second inequality follows from the Rayleigh quotient. By design, $\Rsupt$ becomes tight at high values of $P_r$, while $\Rsupo$ is tighter if $P_r$ is small.

\subsection{Lower Bounds on $\Rs$}

In order to find rates that are actually achievable, it is possible to choose $\F$ as in the derivation of the upper bounds, although these choices will not be optimal in general. One possibility is to choose $\F$ as in \reff{eq:F_ub1}. Another approach, which is derived from the second upper bound $\Rsupt$, is to set
\begin{equation}
\label{eq:F_lowerBound}
\F = \gamma \cdot \frac{\h}{\norm{\h}} \cdot \vmax^H(\R+\W),
\end{equation} 
where $\gamma \in \mathbb{R}$ is chosen such that the relay power constraint \reff{eq:PowerConstr} is fulfilled with equality. Throughout all numerical simulations that have been made, the second approach \reff{eq:F_lowerBound} turned out to deliver better results. For this reason, it will be the only considered lower bound in the remainder of this work. If $\F$ is chosen as in \reff{eq:F_lowerBound}, the achievable rate can be written as
\begin{equation}\label{eq:Rs_sup}
\Rs = \log_2 \of{1 + s + \frac{\norm{\h}^2 \lmax(\R+\W)\gamma^2}{1 + \norm{\h}^2\gamma^2}}.
\end{equation}

\section{TDMA-Based Relaying}\label{sec:TDMA}

In this section, we will introduce a relaying scheme based on TDMA as in \cite{KMH12}. This scheme includes a division of the transmission in $K$ time slots, where  user $k$ occupies the $k$-th time slot exclusively. Also the relay incorporates this slot structure, i.e., the relaying matrix $\Fk$ in time slot $k$ can be adapted to the channel of user $k$ only. Thus, the TDMA slot structure decomposes the channel in $K$ independent single-user relay channels. Therefore, we will first derive the optimal structure of the transmit covariance- and relaying matrix for the single-user relay channel in Subsection \ref{subsec:single_relay}. In the following Subsection \ref{subsec:transmission}, we will transfer this scheme to the MARC with TDMA and derive an algorithm that finds the optimal duration of the time slots, such that the sum-rate is maximized. Finally, in Subsection \ref{subsec:evaluation} we will compare this sum-rate to those derived in section \ref{sec:joint_relaying} for the case $P_r \rightarrow \infty$.

\subsection{Single-User Relaying}\label{subsec:single_relay}

In order to describe a single-user relay channel with a consistent notation, we assume the same channel model as introduced in Subsection \ref{subsec:channel_mod} with only $K=1$ transmitting user. Thus, also \reff{eq:Rs} is valid and can be used to calculate the (sum) rate $\Ro$ of the only user. In contrast to the previous section the optimization of $\Fo$ is strongly simplified as we have $\R = \hro\Po{\hro}^H$, $s=\norm{\hdo}^2 \Po$, and especially $\W = 0$. Hence, the rate $\Ro$ can be written as
\begin{equation}
\label{eq:Ro_tdma}
\Ro = \log_2 \of{1 + s + \frac{\hh\Fo\R{\Fo}^H\h}{1+\hh\Fo{\Fo}^H\h}}.
\end{equation}
The optimization of \reff{eq:Ro_tdma} over $\Fo$ is basically the same as in \reff{eq:Rsup1_pre}. Thus, in analogy to \reff{eq:F_ub1} the optimal $\Fo$ is given by
\begin{equation*}
\Fo = \sqrt{\frac{P_r}{1 + \norm{\hro}^2\Po}}\cdot \frac{\h\cdot{\hro}^H}{\norm{\h}\norm{\hro}},
\end{equation*}
which leads to a rate of
\begin{equation*}
\Ro = \log_2\!\!\of{\!1 \!+\! \norm{\hdo}^2\!\!\Po \!+\! \frac{\norm{\h}^2\norm{\hro}^2\Po P_r}{1 \!+\! \norm{\h}^2 P_r \!+\! \norm{\hro}^2\Po}\!}\!\!.
\end{equation*}

\subsection{TDMA-based Transmission Scheme}\label{subsec:transmission}
The $K$-user MARC is decomposed in $K$ single-user relay channels by using a TDMA scheme, such that user $k$ transmits only in a time slot of duration $\tauk \geq 0$ with $\sum_{k=1}^{K}\tauk = 1$. In each time slot, the optimal choice of the relay matrix $\Fk$  can be obtained as in subsection \ref{subsec:single_relay}. The only difference is the transmit power constraint: As user $k$ only transmits in $\tauk$ fraction of the time, it can use a transmit power of $\Pk/\tauk$ and still fulfills the average transmit power constraint \reff{eq:PowerConstr}. Thus, the rate of user $k$ is given by
\begin{multline}\label{eq:Rk_tdma}
\Rk = \tauk \log_2 \left(1 + \frac{\norm{\hkd}^2\Pk}{\tauk} \right.\\ +\left.\frac{\norm{\h}^2\Pk\norm{\hkr}^2P_r}{\norm{\h}^2P_r\tauk + \Pk\norm{\hkr}^2+\tauk}  \right),
\end{multline}
and the sum-rate can be calculated as $\RsT = \sum_{k=1}^K \Rk$. This sum-rate can be optimized by the choice of $\tauks$, i.e., we are facing the optimization problem
\begin{align}
\max_{\tauks} \;\;&\sum_{k=1}^K \Rk\of{\tauk}\label{eq:opt_tdma}\\
\text{s.t.}\;\;&h(\btau) = 1 - \sum_{k=1}^K \tauk = 0,\notag
\end{align}
where $\btau = \left[ \tauks \right]$. It is easy to see that $\frac{\partial\Rk}{\partial\tauj} = 0$ $\forall j \neq k$ and $\frac{\partial^2\Rk}{\partial{\tauk}^2} < 0$, which makes the problem convex. Thus, the famous Karush-Kuhn-Tucker (KKT) conditions provide necessary and sufficient conditions for optimality. For the above problem, the KKT conditions of a solution $\btaus$ to be optimal can be formulated as $h(\btaus) = 0$ and $\nabla\RsT(\btaus) + \nus \nabla h(\btaus) = 0$, where $\nus \in \mathbb{R}$ can be chosen arbitrarily. As the derivatives of $h$ are directly obtained as $\frac{\partial h}{\partial\tauk} = -1$, the second KKT condition can be rewritten as
\begin{equation*}
\left.\frac{\partial\Rk} {\partial\tauk}\right|_{\tauk=\taukstar} = \nus\;\;\;\forall k,
\end{equation*}
i.e., the derivatives of the individual rates $\Rk$ have to be the same for each user. Due to their lengthiness, those derivatives are not stated here. However, they are straightforward to calculate and it is easy to see that $\left.\frac{\partial\Rk} {\partial\tauk}\right|_{\tauk=0} \rightarrow \infty$ and $\left.\frac{\partial\Rk} {\partial\tauk}\right|_{\tauk \rightarrow \infty} = 0$. Unfortunately, a closed form solution how to optimally choose $\tauks$ as in \cite{KMH12} seems intractable. Therefore, we describe in Algorithm \ref{alg:opt_tau} how the optimal solution can be found iteratively.

\begin{algorithm}
\caption{Iterative optimization of $\tauks$}
\label{alg:opt_tau}
\begin{algorithmic}[1]
\State Set $\taukstar = \frac{1}{K}$ $\forall k=1,\ldots,K$
\While{true}
\State $i = \arg\min_k \;\;\left.\frac{\partial\Rk} {\partial\tauk} \right|_{\tauk = \taukstar}$
\State $j = \arg\max_k \;\;\left.\frac{\partial\Rk} {\partial\tauk} \right|_{\tauk = \taukstar}$
\If{$\left.\frac{\partial\Rj} {\partial\tauj} \right|_{\tauj = \taujstar} - \left.\frac{\partial\Ri} {\partial\taui} \right|_{\taui = \tauistar} > \varepsilon$}
\State Find $0 < \delta < \min\{\taui,1-\tauj\}$, such that
$$\left.\frac{\partial\Rj} {\partial\tauj} \right|_{\tauj = \taujstar + \delta} = \left.\frac{\partial\Ri} {\partial\taui} \right|_{\taui = \tauistar - \delta}$$
\State $\taujstar = \taujstar + \delta$
\State $\tauistar = \tauistar - \delta$
\Else
  \State {\bf break}
\EndIf
\EndWhile
\end{algorithmic}
\end{algorithm}

The main idea of the algorithm is to iteratively equalize the derivatives of $\Rk$ by changing the lengths of the time slots $\tauks$. Therefore, the users $i$ and $j$ with the smallest and largest derivative are selected. Their derivatives are equalized by numerically finding a value $\delta$, which is added to $\tauj$ and subtracted from $\taui$. Due to the properties of the functions discussed above this value can always be found in the interval $(0,\min\{\taui,1-\tauj\})$, and the other derivatives remain unchanged. This procedure is repeated iteratively until the difference of the largest and smallest derivative is at most $\varepsilon$, which can be selected very small to approximate the optimal solution as good as desired.

\subsection{Comparison with Joint Relaying}\label{subsec:evaluation}
As a closed form of the optimum sum-rate of the TDMA scheme has not been found, and for joint relaying only upper and lower bounds are available, comparing these two schemes in an analytical way is not as straightforward as in \cite{KMH12}. Therefore, the two schemes will mainly be compared by the mean of simulation results in Section \ref{sec:simres}. However, an analytic comparison is possible for the asymptotic case of $P_r \rightarrow \infty$, which is described in the following theorem.
\begin{theo}\label{theo:tdma_sup}
In the considered $K$-user MARC with direct links and $P_r \rightarrow \infty$, the joint relaying scheme achieves higher sum-rates than the TDMA-based transmission scheme with optimal time slot durations $\tauk$, if and only if
\begin{equation}\label{ineq:tdma_sup}
 \lmax(\R + \W) > \sum\limits_{k=1}^{K}\norm{\hkr}\Pk.
\end{equation}
\end{theo}
\begin{IEEEproof}
For the joint relaying scheme with $P_r \rightarrow \infty$, we have $\gamma \rightarrow \infty$ in \reff{eq:Rs_sup}, i.e., the sum-rate $\Rs$ converges to $\Rsupt$ in \reff{eq:Rsup2}. Considering the TDMA scheme for $P_r \rightarrow \infty$, the individual user rates $\Rk$ from \reff{eq:Rk_tdma} tend to
\begin{equation*}
\Rkinf = \tauk \log_2 \left[1 + \frac{\Pk}{\tauk} \left( \norm{\hkd}^2 + \norm{\hkr}^2 \right)\right].
\end{equation*}
If this term is used instead of $\Rk$ in the optimization problem \reff{eq:opt_tdma}, it is straightforward to show that choosing $\tauk$ as
\begin{equation*}
\tauk = \frac{\Pk \left( \norm{\hkd}^2 + \norm{\hkr}^2 \right)}{\sum\limits_{j=1}^{K} \Pj \left( \norm{\hjd}^2 + \norm{\hjr}^2 \right)}
\end{equation*}
is the global optimal solution that leads to the sum-rate
\begin{equation}\label{eq:Rsopt_tdma}
\RsTinf = \log_2 \left[1 + \sum\limits_{k=1}^{K} \Pk \left( \norm{\hkd}^2 + \norm{\hkr}^2 \right)\right].
\end{equation}
Comparing \reff{eq:Rsopt_tdma} and \reff{eq:Rsup2}, the theorem follows.
\end{IEEEproof}
Considering especially the matrix $\W$ in \reff{ineq:tdma_sup}, it turns out that, if the direct links $\hkd$ are zero, we have $\W = 0$. Thus, the left hand side in \reff{ineq:tdma_sup} is smaller and TDMA is always better in this case (cf. \cite{KMH12}). However, if the direct links are increased, also $\lmax(\R+\W)$ increases rapidly, which compensates the disadvantage of joint relaying.

\section{Simulation Results}\label{sec:simres}
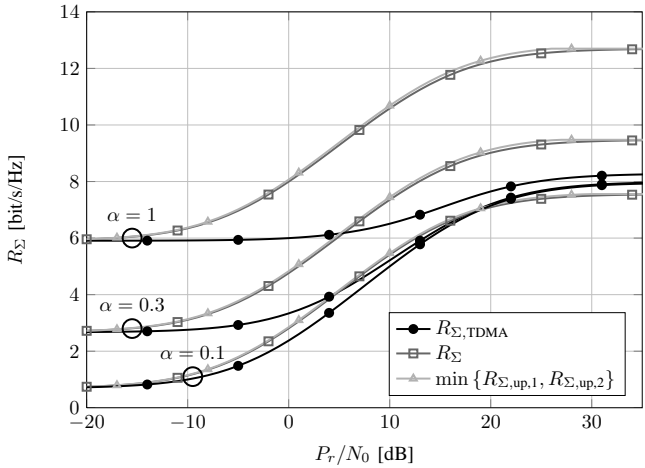
\begin{figure}
\centering{% This file was created by matlab2tikz v0.1.3.
% Copyright (c) 2008--2011, Nico Schlömer <nico.schloemer@gmail.com>
% All rights reserved.
% 
% The latest updates can be retrieved from
%   http://www.mathworks.com/matlabcentral/fileexchange/22022-matlab2tikz
% where you can also make suggestions and rate matlab2tikz.
% 
\begin{tikzpicture}[scale=0.75]
\definecolor{mycolor1}{rgb}{0.4,0.4,0.4}
\definecolor{mycolor2}{rgb}{0.7,0.7,0.7}

\begin{axis}[%
scale only axis,
width=1.1\linewidth,
height=7cm,
xmin=-20, xmax=35,
ymin=0, ymax=14,
xlabel={$P_r/N_0$ [dB]},
ylabel={$\Rs$ [bit/s/Hz]},
xmajorgrids,
ymajorgrids,
zmajorgrids,
legend entries={$\RsT$,$\Rs$,{$\min\left\{\Rsupo,\Rsupt\right\}$}},
legend style={nodes=right},
legend pos=south east
]

\addplot [
color=black,
solid,
mark=*,
mark options={color=black},
mark repeat=9,
mark phase=7,
line width=1.0pt
]
coordinates{
 (-20,2.67753)(-19,2.67992)(-18,2.68293)(-17,2.6867)(-16,2.69142)(-15,2.69734)(-14,2.70474)(-13,2.71398)(-12,2.72548)(-11,2.73978)(-10,2.7575)(-9,2.77937)(-8,2.80626)(-7,2.83913)(-6,2.87909)(-5,2.92733)(-4,2.98512)(-3,3.05375)(-2,3.1345)(-1,3.22856)(0,3.33699)(1,3.46068)(2,3.60023)(3,3.75589)(4,3.92743)(5,4.11412)(6,4.31466)(7,4.52718)(8,4.74939)(9,4.97863)(10,5.21207)(11,5.44678)(12,5.67984)(13,5.90844)(14,6.12991)(15,6.34183)(16,6.54208)(17,6.72892)(18,6.90101)(19,7.0575)(20,7.19798)(21,7.32253)(22,7.43163)(23,7.52609)(24,7.60699)(25,7.67558)(26,7.7332)(27,7.7812)(28,7.82089)(29,7.85349)(30,7.88013)(31,7.90178)(32,7.91931)(33,7.93345)(34,7.94482)(35,7.95395)(36,7.96126)(37,7.96711)(38,7.97178)(39,7.9755)(40,7.97847) 
};

\addplot [
color=mycolor1,
solid,
mark=square,
mark options={color=mycolor1},
mark repeat=9,
line width=1.0pt
]
coordinates{
 (-20,0.742785)(-19,0.756251)(-18,0.772986)(-17,0.793719)(-16,0.819311)(-15,0.850762)(-14,0.889212)(-13,0.935934)(-12,0.992305)(-11,1.05977)(-10,1.1398)(-9,1.23379)(-8,1.34299)(-7,1.46842)(-6,1.61079)(-5,1.77041)(-4,1.94718)(-3,2.14057)(-2,2.34963)(-1,2.57305)(0,2.8092)(1,3.05624)(2,3.31214)(3,3.57476)(4,3.84189)(5,4.11131)(6,4.38075)(7,4.64795)(8,4.91067)(9,5.1667)(10,5.41392)(11,5.6503)(12,5.87401)(13,6.08345)(14,6.27733)(15,6.45475)(16,6.61519)(17,6.75856)(18,6.88519)(19,6.99575)(20,7.09123)(21,7.17283)(22,7.24189)(23,7.29983)(24,7.34804)(25,7.38787)(26,7.42057)(27,7.44726)(28,7.46896)(29,7.48651)(30,7.50067)(31,7.51206)(32,7.5212)(33,7.52851)(34,7.53436)(35,7.53904)(36,7.54276)(37,7.54573)(38,7.5481)(39,7.54998)(40,7.55148) 
};

\addplot [
color=mycolor2,
solid,
mark=triangle,
mark options={color=mycolor2},
mark phase=4,
mark repeat=9,
line width=1.0pt
]
coordinates{
 (-20,0.744692)(-19,0.758626)(-18,0.775935)(-17,0.797368)(-16,0.82381)(-15,0.856284)(-14,0.895952)(-13,0.944107)(-12,1.00214)(-11,1.07152)(-10,1.15371)(-9,1.25007)(-8,1.36186)(-7,1.49004)(-6,1.63526)(-5,1.79779)(-4,1.97746)(-3,2.17368)(-2,2.38544)(-1,2.61139)(0,2.84987)(1,3.09902)(2,3.3568)(3,3.62109)(4,3.88969)(5,4.16039)(6,4.43094)(7,4.69913)(8,4.96272)(9,5.21954)(10,5.46748)(11,5.70453)(12,5.92886)(13,6.1389)(14,6.33335)(15,6.5113)(16,6.67225)(17,6.8161)(18,6.94318)(19,7.05415)(20,7.14999)(21,7.23192)(22,7.30128)(23,7.35946)(24,7.40789)(25,7.4479)(26,7.48076)(27,7.50758)(28,7.52938)(29,7.54702)(30,7.55729)(31,7.55729)(32,7.55729)(33,7.55729)(34,7.55729)(35,7.55729)(36,7.55729)(37,7.55729)(38,7.55729)(39,7.55729)(40,7.55729) 
};

\addplot [
color=mycolor1,
solid,
mark=square,
mark options={color=mycolor1},
mark repeat=9,
line width=1.0pt
]
coordinates{
 (-20,2.72059)(-19,2.73377)(-18,2.75015)(-17,2.77045)(-16,2.79551)(-15,2.82633)(-14,2.86402)(-13,2.90984)(-12,2.96517)(-11,3.03143)(-10,3.1101)(-9,3.20257)(-8,3.31012)(-7,3.43378)(-6,3.57428)(-5,3.73199)(-4,3.90683)(-3,4.09831)(-2,4.30551)(-1,4.52715)(0,4.76163)(1,5.00709)(2,5.26152)(3,5.52277)(4,5.78861)(5,6.05679)(6,6.32502)(7,6.59103)(8,6.85255)(9,7.10735)(10,7.35328)(11,7.58834)(12,7.81066)(13,8.01868)(14,8.21112)(15,8.38708)(16,8.54608)(17,8.68806)(18,8.81337)(19,8.92269)(20,9.01703)(21,9.0976)(22,9.16576)(23,9.2229)(24,9.27043)(25,9.30967)(26,9.34188)(27,9.36817)(28,9.38952)(29,9.4068)(30,9.42073)(31,9.43193)(32,9.44092)(33,9.44812)(34,9.45387)(35,9.45847)(36,9.46213)(37,9.46505)(38,9.46738)(39,9.46923)(40,9.4707) 
};

\addplot [
color=mycolor1,
solid,
mark=square,
mark options={color=mycolor1},
mark repeat=9,
line width=1.0pt
]
coordinates{
 (-20,5.95831)(-19,5.97145)(-18,5.98778)(-17,6.00801)(-16,6.03299)(-15,6.06371)(-14,6.10129)(-13,6.14698)(-12,6.20215)(-11,6.26823)(-10,6.34669)(-9,6.43893)(-8,6.54622)(-7,6.6696)(-6,6.80981)(-5,6.9672)(-4,7.14172)(-3,7.33288)(-2,7.53976)(-1,7.76109)(0,7.99525)(1,8.24041)(2,8.49454)(3,8.75548)(4,9.02102)(5,9.28888)(6,9.55678)(7,9.82243)(8,10.0836)(9,10.3379)(10,10.5834)(11,10.818)(12,11.0398)(13,11.2473)(14,11.4392)(15,11.6146)(16,11.773)(17,11.9145)(18,12.0393)(19,12.1481)(20,12.242)(21,12.3222)(22,12.39)(23,12.4468)(24,12.4941)(25,12.5331)(26,12.5651)(27,12.5912)(28,12.6125)(29,12.6296)(30,12.6435)(31,12.6546)(32,12.6635)(33,12.6707)(34,12.6764)(35,12.681)(36,12.6846)(37,12.6875)(38,12.6898)(39,12.6916)(40,12.6931) 
};

\addplot [
color=black,
solid,
mark=*,
mark options={color=black},
mark repeat=9,
mark phase=7,
line width=1.0pt
]
coordinates{
 (-20,0.724227)(-19,0.733109)(-18,0.744194)(-17,0.757996)(-16,0.77514)(-15,0.796365)(-14,0.822543)(-13,0.854686)(-12,0.893942)(-11,0.941587)(-10,0.999002)(-9,1.06764)(-8,1.14894)(-7,1.24433)(-6,1.35507)(-5,1.48221)(-4,1.62649)(-3,1.78827)(-2,1.96751)(-1,2.1637)(0,2.37592)(1,2.60285)(2,2.84291)(3,3.09425)(4,3.35493)(5,3.62291)(6,3.89609)(7,4.17241)(8,4.44979)(9,4.72613)(10,4.99937)(11,5.26742)(12,5.52824)(13,5.77979)(14,6.02017)(15,6.24759)(16,6.4605)(17,6.65763)(18,6.83807)(19,7.0013)(20,7.14722)(21,7.27614)(22,7.38874)(23,7.486)(24,7.56914)(25,7.63952)(26,7.69856)(27,7.7477)(28,7.7883)(29,7.82162)(30,7.84883)(31,7.87094)(32,7.88883)(33,7.90326)(34,7.91486)(35,7.92417)(36,7.93162)(37,7.93758)(38,7.94234)(39,7.94614)(40,7.94917) 
};

\addplot [
color=black,
solid,
mark=*,
mark options={color=black},
mark repeat=9,
mark phase=7,
line width=1.0pt
]
coordinates{
 (-20,5.90714)(-19,5.9074)(-18,5.90773)(-17,5.90814)(-16,5.90866)(-15,5.90931)(-14,5.91013)(-13,5.91116)(-12,5.91246)(-11,5.91408)(-10,5.91612)(-9,5.91868)(-8,5.92188)(-7,5.92589)(-6,5.93089)(-5,5.93712)(-4,5.94487)(-3,5.95447)(-2,5.96633)(-1,5.98091)(0,5.99875)(1,6.02045)(2,6.04669)(3,6.07817)(4,6.11565)(5,6.15987)(6,6.2116)(7,6.27155)(8,6.34042)(9,6.4188)(10,6.50698)(11,6.60483)(12,6.7117)(13,6.82627)(14,6.94661)(15,7.07032)(16,7.19471)(17,7.31709)(18,7.43497)(19,7.54624)(20,7.64929)(21,7.74305)(22,7.82696)(23,7.90091)(24,7.9652)(25,8.02038)(26,8.0672)(27,8.10655)(28,8.13931)(29,8.16638)(30,8.18861)(31,8.20674)(32,8.22147)(33,8.23338)(34,8.24298)(35,8.2507)(36,8.2569)(37,8.26185)(38,8.26581)(39,8.26898)(40,8.2715) 
};

\addplot [
color=mycolor2,
solid,
mark=triangle,
mark options={color=mycolor2},
mark phase=4,
mark repeat=9,
line width=1.0pt
]
coordinates{
 (-20,2.72367)(-19,2.73761)(-18,2.75492)(-17,2.77635)(-16,2.80279)(-15,2.83527)(-14,2.87493)(-13,2.92309)(-12,2.98113)(-11,3.05051)(-10,3.13269)(-9,3.22906)(-8,3.34084)(-7,3.46902)(-6,3.61424)(-5,3.77677)(-4,3.95645)(-3,4.15266)(-2,4.36442)(-1,4.59037)(0,4.82886)(1,5.078)(2,5.33579)(3,5.60007)(4,5.86867)(5,6.13937)(6,6.40992)(7,6.67811)(8,6.9417)(9,7.19852)(10,7.44646)(11,7.68351)(12,7.90784)(13,8.11788)(14,8.31233)(15,8.49028)(16,8.65123)(17,8.79508)(18,8.92216)(19,9.03313)(20,9.12898)(21,9.2109)(22,9.28026)(23,9.33845)(24,9.38687)(25,9.42689)(26,9.45974)(27,9.47641)(28,9.47641)(29,9.47641)(30,9.47641)(31,9.47641)(32,9.47641)(33,9.47641)(34,9.47641)(35,9.47641)(36,9.47641)(37,9.47641)(38,9.47641)(39,9.47641)(40,9.47641) 
};

\addplot [
color=mycolor2,
solid,
mark=triangle,
mark options={color=mycolor2},
mark phase=4,
mark repeat=9,
line width=1.0pt
]
coordinates{
 (-20,5.96157)(-19,5.97551)(-18,5.99282)(-17,6.01425)(-16,6.04069)(-15,6.07316)(-14,6.11283)(-13,6.16099)(-12,6.21903)(-11,6.2884)(-10,6.37059)(-9,6.46695)(-8,6.57874)(-7,6.70692)(-6,6.85214)(-5,7.01467)(-4,7.19434)(-3,7.39056)(-2,7.60232)(-1,7.82827)(0,8.06676)(1,8.3159)(2,8.57368)(3,8.83797)(4,9.10657)(5,9.37727)(6,9.64782)(7,9.91601)(8,10.1796)(9,10.4364)(10,10.6844)(11,10.9214)(12,11.1457)(13,11.3558)(14,11.5502)(15,11.7282)(16,11.8891)(17,12.033)(18,12.1601)(19,12.271)(20,12.3669)(21,12.4488)(22,12.5182)(23,12.5763)(24,12.6248)(25,12.6648)(26,12.6976)(27,12.6988)(28,12.6988)(29,12.6988)(30,12.6988)(31,12.6988)(32,12.6988)(33,12.6988)(34,12.6988)(35,12.6988)(36,12.6988)(37,12.6988)(38,12.6988)(39,12.6988)(40,12.6988) 
};

\node[circle,draw,line width=1.0pt] (first) at (axis cs:-9.5,1.1) {};
\node[circle,draw,line width=1.0pt] (second) at (axis cs:-15.5,2.8) {};
\node[circle,draw,line width=1.0pt] (third) at (axis cs:-15.5,6) {};

\node[above] at (first.north) {$\alpha=0.1$};
\node[above] at (second.north) {$\alpha=0.3$};
\node[above] at (third.north) {$\alpha=1$};

\end{axis}
\end{tikzpicture}}
\caption{Comparison of TDMA-based and joint relaying scheme for $K=10$, $M_r = 4$, $\Pmax/N_0 = 10$ dB}
\label{fig:TDMAvsSP}
\end{figure}

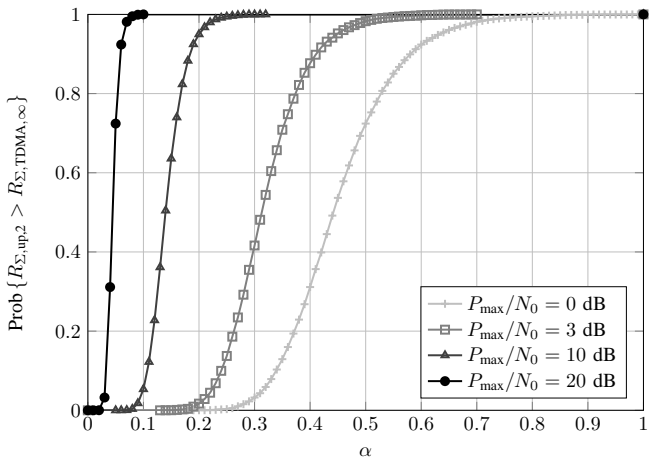
\begin{figure}
\centering{% This file was created by matlab2tikz v0.1.3.
% Copyright (c) 2008--2011, Nico Schlömer <nico.schloemer@gmail.com>
% All rights reserved.
% 
% The latest updates can be retrieved from
%   http://www.mathworks.com/matlabcentral/fileexchange/22022-matlab2tikz
% where you can also make suggestions and rate matlab2tikz.
% 
\begin{tikzpicture}[scale=0.75]

% defining custom colors
\definecolor{mycolor1}{rgb}{0.75,0.75,0.75}
\definecolor{mycolor2}{rgb}{0.5,0.5,0.5}
\definecolor{mycolor3}{rgb}{0.25,0.25,0.25}
\definecolor{mycolor4}{rgb}{0,0,0}

\begin{axis}[%
scale only axis,
width=1.1\linewidth,
height=7cm,
xmin=0, xmax=1,
ymin=0, ymax=1,
xlabel={$\alpha$},
ylabel={$\text{Prob}\left\{ \Rsupt > \RsTinf\right\}$},
xmajorgrids,
ymajorgrids,
zmajorgrids,
legend entries={$\Pmax/N_0 = 0\text{ dB}$,$\Pmax/N_0 = 3\text{ dB}$,$\Pmax/N_0 = 10\text{ dB}$,$\Pmax/N_0 = 20\text{ dB}$},
legend style={nodes=right},
legend pos=south east
]

\addplot [
color=mycolor1,
solid,
line width=1.0pt,
mark=+,
mark options={color=mycolor1},
mark repeat=1
]
coordinates{
 (0,0)(0.18,0)(0.19,0.0001)(0.2,0.0002)(0.21,0.0002)(0.22,0.0004)(0.23,0.0012)(0.24,0.0022)(0.25,0.003)(0.26,0.0055)(0.27,0.0101)(0.28,0.0155)(0.29,0.0226)(0.3,0.0323)(0.31,0.0432)(0.32,0.0589)(0.33,0.0789)(0.34,0.103)(0.35,0.1292)(0.36,0.1598)(0.37,0.1938)(0.38,0.2291)(0.39,0.268)(0.4,0.3113)(0.41,0.3566)(0.42,0.3987)(0.43,0.445)(0.44,0.4918)(0.45,0.5366)(0.46,0.5784)(0.47,0.6185)(0.48,0.6565)(0.49,0.6932)(0.5,0.7242)(0.51,0.7525)(0.52,0.7798)(0.53,0.8068)(0.54,0.8281)(0.55,0.8502)(0.56,0.8673)(0.57,0.8843)(0.58,0.8991)(0.59,0.9122)(0.6,0.9238)(0.61,0.9332)(0.62,0.9404)(0.63,0.9482)(0.64,0.9545)(0.65,0.9608)(0.66,0.9658)(0.67,0.9699)(0.68,0.9737)(0.69,0.9772)(0.7,0.9813)(0.71,0.9837)(0.72,0.9861)(0.73,0.988)(0.74,0.9897)(0.75,0.9909)(0.76,0.9918)(0.77,0.9932)(0.78,0.9936)(0.79,0.9944)(0.8,0.9953)(0.81,0.9958)(0.82,0.9961)(0.83,0.9968)(0.84,0.9972)(0.85,0.9978)(0.86,0.9984)(0.87,0.9986)(0.88,0.9988)(0.89,0.9991)(0.9,0.9992)(0.91,0.9997)(0.92,0.9997)(0.93,0.9997)(0.94,0.9997)(0.95,0.9997)(0.96,0.9997)(0.97,0.9997)(0.98,0.9999)(0.99,1)(1,1)
};

\addplot [
color=mycolor2,
solid,
line width=1.0pt,
mark=square,
mark options={color=mycolor2},
mark repeat=1
]
coordinates{
 (0,0)(0.13,0)(0.14,0.0002)(0.15,0.0002)(0.16,0.001)(0.17,0.0022)(0.18,0.0037)(0.19,0.0093)(0.2,0.0169)(0.21,0.0285)(0.22,0.0444)(0.23,0.0684)(0.24,0.0997)(0.25,0.1373)(0.26,0.186)(0.27,0.2346)(0.28,0.292)(0.29,0.3548)(0.3,0.4163)(0.31,0.4812)(0.32,0.5441)(0.33,0.6042)(0.34,0.657)(0.35,0.7086)(0.36,0.7479)(0.37,0.787)(0.38,0.8218)(0.39,0.8525)(0.4,0.8767)(0.41,0.8981)(0.42,0.9166)(0.43,0.9312)(0.44,0.9412)(0.45,0.952)(0.46,0.9604)(0.47,0.9674)(0.48,0.9729)(0.49,0.9782)(0.5,0.9828)(0.51,0.9861)(0.52,0.9889)(0.53,0.9909)(0.54,0.9921)(0.55,0.9935)(0.56,0.9945)(0.57,0.9956)(0.58,0.9961)(0.59,0.9968)(0.6,0.9977)(0.61,0.9984)(0.62,0.9986)(0.63,0.9991)(0.64,0.9995)(0.65,0.9997)(0.66,0.9997)(0.67,0.9997)(0.68,0.9997)(0.69,0.9999)(0.7,1)(1,1)
};

\addplot [
color=mycolor3,
solid,
line width=1.0pt,
mark=triangle,
mark options={color=mycolor3},
mark repeat=1
]
coordinates{
 (0,0)(0.05,0)(0.06,0.0001)(0.07,0.0005)(0.08,0.0035)(0.09,0.0179)(0.1,0.0534)(0.11,0.1223)(0.12,0.2275)(0.13,0.3613)(0.14,0.5046)(0.15,0.6358)(0.16,0.7399)(0.17,0.8236)(0.18,0.883)(0.19,0.9247)(0.2,0.9498)(0.21,0.9675)(0.22,0.98)(0.23,0.9875)(0.24,0.9916)(0.25,0.9944)(0.26,0.9963)(0.27,0.9979)(0.28,0.999)(0.29,0.9997)(0.3,0.9997)(0.31,0.9999)(0.32,1)(1,1)
};

\addplot [
color=mycolor4,
solid,
line width=1.0pt,
mark=*,
mark options={color=mycolor4},
mark repeat=1
]
coordinates{
 (0,0)(0.01,0)(0.02,0.0002)(0.03,0.0323)(0.04,0.3113)(0.05,0.7242)(0.06,0.9238)(0.07,0.9813)(0.08,0.9953)(0.09,0.9992)(0.1,1)(1,1)
};

\end{axis}
\end{tikzpicture}}
\caption{Probabilities of joint relaying being better than TDMA at $P_r \rightarrow \infty$ for $K=10$, $M_r = 4$}
\label{fig:probs}
\end{figure}

In order to see how the strength of the direct links affects the superiority of either TDMA or joint relaying, we will evaluate the achievable sum-rates by Monte Carlo simulations in this section. Moreover, the average gap between upper bounds and achievable rates shall be investigated.

For this purpose, we assume channels with independent Rayleigh-fading, i.e., all entries of the channel matrices\linebreak are $\sim \mathcal{CN}(0,1)$ and independent from each other. As an exception, the channel gains for the direct links $\hkd$ are multiplied with a parameter $\alpha \in \mathbb{R}$, i.e., they are $\sim \mathcal{CN}(0,\alpha^2)$ in order to vary the strength of the direct links. All results presented in this section are obtained by averaging over at least 1000 channel realizations and the transmit powers $\Pk$ of the users are assumed to be uniformly distributed between $0$ and $\Pmax$.

For a system with $K=10$ users, the sum-rates of both TDMA and joint relaying are plotted in Figure \ref{fig:TDMAvsSP}. We assumed that the relay has $M_r = 4$ antennas and that the maximum transmit power to noise ratio is $\Pmax/N_0 = 10$ dB, while both the strength of the direct channels $\alpha$ and the power at the relay $P_r$ are varied. It can be directly observed that for all parameters the minimum of the two upper bounds and the achievable rates of the joint relaying scheme are very close to each other. Concerning the comparison with TDMA, it can be seen that for $\alpha = 0.1$, the superiority of TDMA discussed in \cite{KMH12} persists only for large values of $P_r$. If the strength of the direct links is further increased, the sum-rates achieved by TDMA grow only slowly, especially for large values of $P_r$. On the other hand, the sum-rates of the joint relaying scheme grow faster, such that they are clearly higher than those of TDMA for $\alpha = 0.3$ and especially for $\alpha=1$.

A further comparison of TDMA and joint relaying for different maximum transmit powers is given in Figure \ref{fig:probs}. In this figure, we plotted the probability that joint relaying achieves higher sum-rates than TDMA at $P_r \rightarrow \infty$, i.e., the probability that \reff{ineq:tdma_sup} holds, for different values of $\alpha$ and $\Pmax$. As already observed in Figure \ref{fig:TDMAvsSP} these probabilities grow with the strength of the direct links $\alpha$. Moreover, it can be noticed that the probabilities also grow with increasing transmit powers. Note that, also if \reff{ineq:tdma_sup} does not hold, joint relaying can still achieve higher rates for lower $P_r$ (cf. plots for $\alpha=0.1$ in Figure \ref{fig:TDMAvsSP}).

\section{Conclusion}\label{sec:conclusion}

We have considered a $K$-user MARC with direct links, where only the relay has multiple antennas. For this channel, we found upper and lower bounds on the achievable sum-rate when all stations transmit their signals jointly. It was shown by Monte Carlo simulations that these bounds can be very close to each other. If a TDMA protocol is used, we were able to determine the optimal relaying matrices $\Fk$ and obtained the optimal time slot durations by an iterative algorithm. For the asymptotic case of $P_r \rightarrow \infty$, an analytic expression was derived, which allows to determine whether joint relaying or TDMA achieve a higher sum-rate. For general values of $P_r$ the performance of joint relaying and TDMA were also found by Monte Carlo simulations. It turned out that the superiority of TDMA found in \cite{KMH12} for absent direct links gets lost quickly if the strength of the direct links is increased. This trend is accelerated if the available transmit powers are increased.

%\section*{Acknowledgment}

%\appendices



%\bibliographystyle{IEEEtran}

\begin{thebibliography}{1}
\providecommand{\url}[1]{#1}
\csname url@samestyle\endcsname
\providecommand{\newblock}{\relax}
\providecommand{\bibinfo}[2]{#2}
\providecommand{\BIBentrySTDinterwordspacing}{\spaceskip=0pt\relax}
\providecommand{\BIBentryALTinterwordstretchfactor}{4}
\providecommand{\BIBentryALTinterwordspacing}{\spaceskip=\fontdimen2\font plus
\BIBentryALTinterwordstretchfactor\fontdimen3\font minus
  \fontdimen4\font\relax}
\providecommand{\BIBforeignlanguage}[2]{{%
\expandafter\ifx\csname l@#1\endcsname\relax
\typeout{** WARNING: IEEEtran.bst: No hyphenation pattern has been}%
\typeout{** loaded for the language `#1'. Using the pattern for}%
\typeout{** the default language instead.}%
\else
\language=\csname l@#1\endcsname
\fi
#2}}
\providecommand{\BIBdecl}{\relax}
\BIBdecl

\bibitem{CG79}
\BIBentryALTinterwordspacing
T.~Cover and A.~E. Gamal, ``{Capacity Theorems for the Relay Channel},''
  \emph{Information Theory, IEEE Transactions on}, vol.~25, no.~5, pp.
  572--584, Sep. 1979. [Online]. Available:
  \url{http://ieeexplore.ieee.org/xpls/abs\_all.jsp?arnumber=1056084}
\BIBentrySTDinterwordspacing

\bibitem{FHK06}
\BIBentryALTinterwordspacing
Z.~Fang, Y.~Hua, and J.~C. Koshy, ``{Joint Source and Relay Optimization for a
  Non-Regenerative MIMO Relay},'' in \emph{Sensor Array and Multichannel
  Processing, 2006. Fourth IEEE Workshop on}.\hskip 1em plus 0.5em minus
  0.4em\relax IEEE, Jul. 2006, pp. 239--243. [Online]. Available:
  \url{http://dx.doi.org/10.1109/SAM.2006.1706129}
\BIBentrySTDinterwordspacing

\bibitem{YH10}
\BIBentryALTinterwordspacing
Y.~Yu and Y.~Hua, ``{Power Allocation for a MIMO Relay System With
  Multiple-Antenna Users},'' \emph{IEEE Transactions on Signal Processing},
  vol.~58, no.~5, pp. 2823--2835, May 2010. [Online]. Available:
  \url{http://dx.doi.org/10.1109/TSP.2010.2042476}
\BIBentrySTDinterwordspacing

\bibitem{KMH12}
\BIBentryALTinterwordspacing
F.~Knabe, O.~Mohamed, and C.~Huppert, ``{Superiority of TDMA in a Class of
  Gaussian Multiple-Access Channels with a MIMO-AF-Relay},''
  \emph{arXiv:1202.1734v2}, Mar. 2012. [Online]. Available:
  \url{http://arxiv.org/abs/1202.1734}
\BIBentrySTDinterwordspacing

\bibitem{TH07}
\BIBentryALTinterwordspacing
X.~Tang and Y.~Hua, ``{Optimal Design of Non-Regenerative MIMO Wireless
  Relays},'' \emph{Wireless Communications, IEEE Transactions on}, vol.~6,
  no.~4, pp. 1398--1407, Apr. 2007. [Online]. Available:
  \url{http://dx.doi.org/10.1109/TWC.2007.348336}
\BIBentrySTDinterwordspacing

\bibitem{VSBS06}
\BIBentryALTinterwordspacing
N.~Varanese, O.~Simeone, Y.~Bar-Ness, and U.~Spagnolini, ``{Achievable Rates of
  Multi-Hop and Cooperative MIMO Amplify-and-Forward Relay Systems with Full
  CSI},'' in \emph{Signal Processing Advances in Wireless Communications, 2006.
  SPAWC '06. IEEE 7th Workshop on}.\hskip 1em plus 0.5em minus 0.4em\relax
  IEEE, Jul. 2006, pp. 1--5. [Online]. Available:
  \url{http://dx.doi.org/10.1109/SPAWC.2006.346423}
\BIBentrySTDinterwordspacing

\end{thebibliography}

\bibliographystyle{IEEEtran}


\end{document}